\documentclass[preprint,onecolumn,amsmath,amssymb]{revtex4}
\usepackage[pdftex]{graphicx}

\newcommand{\remark}[1]{}

\newcommand{\ket}[1]{\ensuremath{|#1\rangle}}

\newcommand{\Ca}{$^{40}$Ca$^+\,$}

\begin{document}

\title[Wiring up trapped ions]{Wiring up trapped ions to study aspects of quantum information}

\author{N.~Daniilidis}
    \affiliation{Institut f\"ur Quantenoptik and Quanteninformation, Innsbruck, Austria}
\author{T.~Lee}
    \altaffiliation{Dept. of Physics, California Institute of Technology, Pasadena, CA, USA}
\author{R.~Clark}
    \altaffiliation{Center for Ultracold Atoms, Massachusetts Institute of Technology, Cambridge, MA, USA}
\author{S.~Narayanan}
    \affiliation{Institut f\"ur Quantenoptik and Quanteninformation, Innsbruck, Austria}
\author{H.~H\"affner}
    \altaffiliation{Dept. of Physics, University of California, Berkeley, CA 94720, USA}
    \altaffiliation{Materials Sciences Division, Lawrence Berkeley National Laboratory, Berkeley, CA 94720, USA}
    \email{hhaeffner@berkeley.edu}

\begin{abstract}
There has been much interest in developing methods for transferring quantum information. We discuss a way to transfer quantum information between two trapped ions through a wire. The motion of a trapped ion induces oscillating charges in the trap electrodes. By sending this current to the electrodes of a nearby second trap, the motions of ions in the two traps are coupled. We investigate the electrostatics of a set-up where two separately trapped ions are coupled through an electrically floating wire. We also discuss experimental issues, including possible sources of decoherence.
\end{abstract}

\maketitle

\section{Introduction}
Trapped ions are an ideal system to store and process quantum information. Held by
electrodynamical forces inside a vacuum chamber, the ions hardly interact with the environment.
Thus, extremely long coherence times of up to 10~minutes have been demonstrated with trapped
ions \cite{Bollinger1991,Fisk:1997}. Furthermore, the internal and external degrees of freedom
can be controlled very accurately with laser radiation \cite{Leibfried2003b}. With
these techniques, researchers have implemented quantum gates \cite{Leibfried2003a,Schmidt-Kaler2003},
multiparticle entanglement \cite{Sackett2000,Leibfried2005,Haeffner2005a}, and even basic
quantum algorithms \cite{Riebe2004,Barrett2004,Chiaverini:2004a}. For a review of \mbox{ion--trap}
quantum computation see \cite{Haeffner:2008}. Currently, the goal is to expand to larger
quantum registers in order to be able to use quantum computing to solve nontrivial problems. One strategy is
to work with small and easy to control ion strings and then physically transport ions between different zones
\cite{Kielpinski2001}. Other schemes transfer quantum information via
optical cavities \cite{Cirac1997,Kimble:2008} and long-distance entanglement \cite{Moehring2007,Gottesman1999}.

In this contribution, we concentrate on a
different coupling mechanism: ions in two separate traps can be coupled by allowing the charges they induce in the electrodes to affect each other's motion \cite{Wineland1975,Heinzen1990}. This inter--trap coupling may be used for
scalable quantum computing, cooling ion species that cannot be laser cooled, and for coupling an \mbox{ion--trap} quantum computer to a \mbox{solid--state} quantum computer, e.g. a system of Josephson junctions \cite{Tian2004a,Tian2005}. Related schemes have been proposed for coupling Rydberg atoms \cite{Soerensen2004} and oscillating electrons \cite{Stahl2005}.

\section{Coupling mechanism}
\label{sec:coupling}
In the following, we discuss experiments toward inter-trap coupling using \Ca ions in a
planar RF trap. Experiments in a similar direction with electrons trapped in Penning traps are discussed
in Ref.~\cite{Marzoli:2008}. The basic idea is that quantum information stored in the electronic
degree of freedom of a single ion cooled to the motional ground state can be mapped onto the motional
degree of freedom by driving the motional sidebands of the electronic transition \cite{Leibfried2003b}.
Thus, the information is stored in superpositions of the form $\alpha\ket{0}+\beta\ket{1}$, where $\ket{n}$ is
the quantum number of the harmonic oscillator describing ion motion. This oscillating motion yields
a considerable dipole moment which can be coupled to the motion of an ion in a different trap.
For instance, starting with one ion in $(\ket{0}+\ket{1})/\sqrt{2}$ and the other ion in $\ket{0}$, we expect that after some time $t_{\rm ex}$, the ions have exchanged states to within acquired phases. The main idea of these experiments is to enhance the coupling using a wire and thus provide a valuable means of interconnecting trapped ions.

\begin{figure}[t!]
\begin{center}
\includegraphics[width=0.6\textwidth]{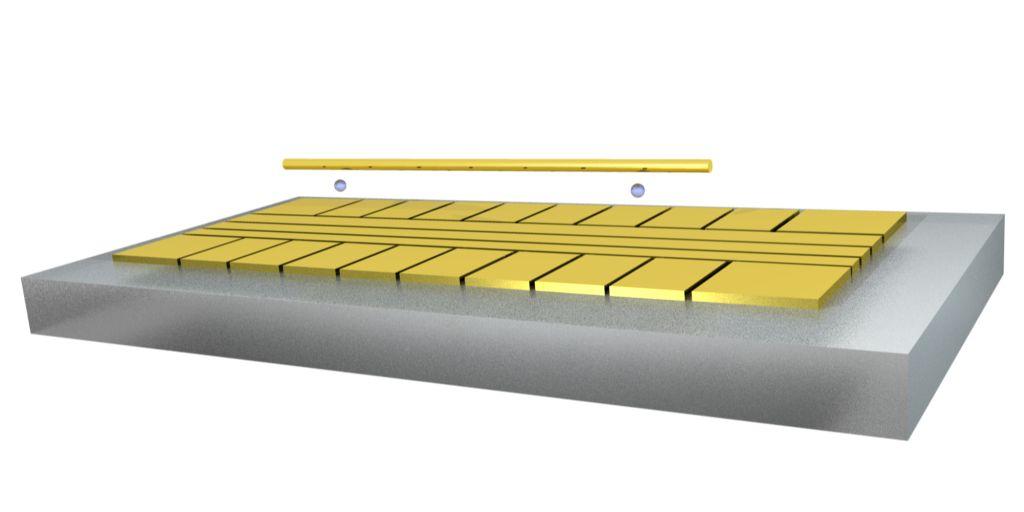}
\end{center}
\caption{\label{fig:set-up}
Schematic representation of the experimental setup used for the \mbox{trapped--ion} coupling experiments. A planar trap
with several DC electrode segments provides multiple trapping regions on the same trap chip. An
electrically floating electrode is in proximity to ions in different trapping regions and couples their motional
states.
}
\end{figure}

Fig.~\ref{fig:set-up} sketches the experimental setup. A planar RF trap confines two ions
in two different potential wells above the trap surface. An electrically floating wire mounted above the trap
carries currents induced by ion motion, enhancing the Coulombic coupling between ions. In what follows, we study the  dynamics of the coupled ion system using the system Hamiltonian and also using the equivalent circuit approach, described in Refs~\cite{Wineland1975,Heinzen1990}.

\subsection{Ion-wire interaction}

\begin{figure}[!tb]
\centering
\includegraphics[width=0.35\textwidth]{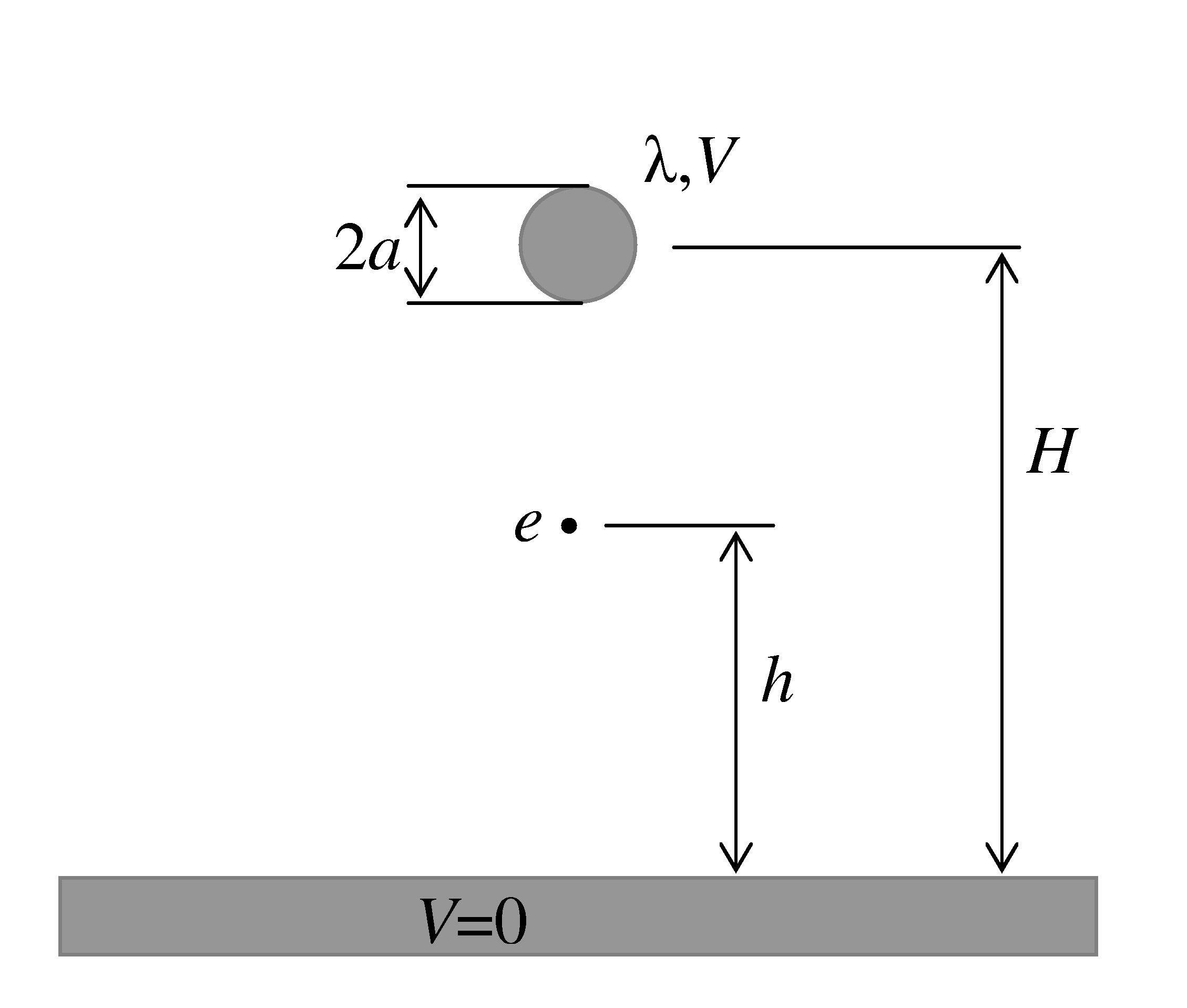}
\caption{ Schematic of a wire of radius $a$, length $L$, and total charge $\lambda L$ at height $H$ and the ion at height $h$ above the grounded plane. The wire is coming out of the page.}
\label{fig:plane-ion-wire-setup}
\end{figure}

We first derive the electrostatic coupling term, under some simplifying assumptions. Our analysis
follows a procedure similar to that described in \cite{Shockley:1938}. We consider a wire of radius $a$ and length
$L$, situated some height $H$ above a (infinite) ground plane and oriented parallel to the plane (see Fig.~\ref{fig:plane-ion-wire-setup}). Two point charges, henceforth ion \#1 and ion \#2, are at heights
$h_1,\,h_2$ ($h_1,\,h_2<H$) above the ground plane, located on the plane passing through the center of
the wire and vertical to the ground plane. The horizontal distance, $d$, between the ions satisfies
$h_1,\,h_2,\,H\ll d<L$. The point charges are treated here as infinitesimally small conductors with
variable, externally set, charge. Consider the situation where the wire is at potential $V$ and carries
a total charge $\lambda\cdot L$, while the ``point charge'' conductors carry zero charge. Then,
in the limit $ L,\,d\,\gg\,H,\,a$, the potentials $V$ of the wire, and $\Phi_{1,2}$ at the ion positions are
\begin{equation}\label{eq:wire-potential}
 V=\frac{\lambda}{2\,\pi\,\epsilon_0}\,\ln\left( \frac{2\,H-a}{a}\right)\,,\;\;
 \Phi_{i}=\frac{\lambda}{2\,\pi\,\epsilon_0}\,\ln\left( \frac{H+h_{ i}}{H-h_{ i}}\right),\; i=1,\,2.
\end{equation}
A convenient dual situation is the one in which both point charges carry the same charge, $e$, while the
wire carries zero net charge, and is at potential $V'$. Application of Green's theorem to the above situations
results in the relation:
\begin{equation}\label{eq:induced-wire-potential}
 V'=\frac{e}{2\,\pi\,\epsilon_0\,L}\,\left[\ln\left( \frac{H+h_1}{H-h_1}\right)+\ln\left( \frac{H+h_2}{H-h_2}\right)\right]\;.
\end{equation}
We are interested in the \mbox{ion--ion} interaction that is mediated by the wire, and we neglect
their direct electrostatic interaction. The potential energy of each ion in the wire potential is
\begin{equation}\label{eq:interaction-energy}
 U_{ i} = \frac{e\,V'}{\alpha}\,\ln\left(\frac{H+h_{ i}}{H-h_{ i}}\right),\;i=1,\,2,
\end{equation}
where $ \alpha=\ln\left[ (2\,H-a)/a\right]$. For the remainder of this section, it is convenient to switch to coordinates in
which the instantaneous height of each ion is denoted by the deviation, $y_{ i}$, about the equilibrium height,
$h_{i}=h_{0,i}+y_{ i}$.
In these coordinates, the coupling constant that enters the Hamiltonian of the system is
\begin{equation}\label{eq:couping-constant}
\gamma\equiv \frac{1}{2}\frac{\partial^2 (U_1+U_2)}{\partial y_1 \partial y_2}
= \frac{2\,e^2\,H^2}{\pi\,\epsilon_0\,\alpha\,L}\cdot\frac{1}{(H^2-h_{0,1}^2)(H^2-h_{0,2}^2)}\;.
\end{equation}
Here the factor $1/2$ is introduced to avoid double counting of the electrostatic energy, as described in \cite{Jackson:1998}. As stated above, each ion is confined in an independent harmonic trap. Thus the Hamiltonian
for the coupled ion system in the presence of the floating wire is
\begin{equation}\label{eq:hamiltonian}
H=\frac{p_1^2}{2\,m}+\frac{1}{2}m\,\omega_1^2\,y_1^2+
  \frac{p_2^2}{2\,m}+\frac{1}{2}m\,\omega_2^2\,y_2^2
  +\gamma\,y_1\,y_2\;.
\end{equation}

The time evolution of the above Hamiltonian has been studied for the resonant case
($\omega_1=\omega_2$) exactly and also in the rotating wave approximation \cite{Estes:1968}.
It was found that the rotating wave approximation is in almost complete agreement with the
exact solution in the limit of small coupling constants ($\gamma/m\omega^2<0.1$). More
recently, a solution in the rotating wave approximation showed that full exchange of
motional states occurs only in the resonant case and for specific initial states \cite{Portes:2008}.
One interesting case is with one ion initially in a superposition of Fock states of the form $(\ket{0}+\ket{n})/\sqrt{2}$ and the second ion in the ground state. In this case, the inverse time for
state exchange of the two ions is
\begin{equation}\label{eq:exchange-rate}
\frac{1}{t_{\rm ex}}=\frac{\gamma}{\pi\,\omega\, m}=\frac{2\,e^2\,H^2}{\pi^2\,\epsilon_0\,\alpha\,m\,\omega\, L} \cdot\frac{1}{(H^2-h_{0,1}^2)(H^2-h_{0,2}^2)}\;,
\end{equation}
where the geometry constant $\alpha$ was defined below Eq.~\ref{eq:interaction-energy}. After time $t_{\rm ex}$, the first ion is in the ground state and the second ion is in $(\ket{0}+e^{-in\Theta}\ket{n})/\sqrt{2}$, where $\Theta = \pi(m\omega^2/\gamma+1/2)$. In experiments aiming to transfer quantum information, the presence of the acquired phase $\Theta$ poses requirements similar to those for preserving the coherence of the motional state of a single ion.

Another important case concerns coupling of coherent states in the resonant system. It is easy to verify that if the first ion starts out in a coherent state $\ket{\mu}$, with complex amplitude $\mu$, and the second ion in the ground state, then after time $t_{\rm ex}$ the first ion is in the ground state and the second ion in a coherent state $\ket{\mu\,e^{-i\Theta}}$, where $\Theta$, defined above, describes the change of the coherent state complex amplitude. This effect will be present in the classical regime. It is due to the fact that each oscillator continues to oscillate while the state exchange is in process and thus acquires some phase. The presence of such a phase can be most easily observed by allowing the coupled ions to evolve for time $2\,t_{\rm ex}$, so that the first ion has returned to a state $\ket{\mu\,e^{-i2\Theta}}$.

An important aspect of the above result is the extraction of the dependence of the coupling rate on
experimental parameters. The coupling rate increases with decreasing size of the experimental setup:
the length of the wire as well as the ion--wire distances enter mainly as $1/\left[L\,(H-h_{0,1})\,(H-h_{0,2})\right]$.
Dependence on the wire radius is only logarithmic, included in the geometric constant $\alpha$. Physically,
the increased coupling with smaller system sizes corresponds to the fact that for ions closer to the wire
the induced charges are larger, and also that for shorter wires the induced charges are distributed over
shorter distances. The scaling of $t_{\rm ex}$ with system size yields a decrease of $t_{\rm ex}$ by roughly an order of magnitude for a decrease of the trap size, i.e. $H$, $h_{0,i}$, and wire length, $L$, by a factor of 2.
Besides these geometrical considerations, we find an inverse law dependence of the coupling rate on the ion
secular frequencies, $t_{\rm ex}\propto \omega$. This can be understood physically by the increase in effective coupling as the dipole moment corresponding to each ion ($\propto 1/\sqrt{\omega}$) increases with lower secular frequency.

Typical parameters that are feasible in current setups are: $H = 200\,\rm \mu m$,
$h_{0,i} = 150\,\rm \mu m$, $L = 10\,\rm mm$, $a = 12.5\,\rm\mu m$, $\omega = 2\,\pi\cdot1 \rm MHz$.
With these values and for two \Ca ions, one obtains  $t_{\rm ex}\approx190 \,\rm ms$.

It is interesting to note that the Hamiltonian for the case of $N$ individually trapped ions is
\begin{equation}
H = \sum_{i=1}^{N}H_{\rm{h.o.},\emph i}+\sum_i\sum_{j>i}\gamma_{i,j}y_i y_j\;,
\end{equation}
where $H_{\rm{h.o.},\emph i}$ is the harmonic oscillator Hamiltonian for the $i^{\rm{th}}$ ion. This involves coupling of neighbors to all orders, and can be interesting in the non-trivial case where the coupling constants $\gamma_{i,j}$ vary with neighbor separation.

\subsection{Equivalent circuit model}

We now derive the circuit model for the system of two trapped ions
and a coupling wire, following Ref. \cite{Wineland1975,Heinzen1990}. The analysis starts from the equations of motion for the ions
\begin{equation}\label{eq:equations-of-motion}
\frac{e}{m}E_{ i}=\ddot{y_{ i}}+\omega_{ i}^2 y_{ i}\,,\;i=1,\,2\;.
\end{equation}
Here, the electric field $E_i$ at the position of ion $i$ is due to the potential of the floating
wire used for coupling. The field $E_{ i}$ is proportional to the potential $V_{ i}$ of the wire. In addition,
following the analysis described in \cite{Shockley:1938}, we find that in the presence of a single ion at height $h$,
the charge induced in the wire is
\begin{equation}\label{eq:induced-charge}
 q_{\rm ind}=-\frac{e}{\alpha}\,\ln\left(\frac{H+h}{H-h}\right)\;,
\end{equation}
A key consequence of this result is that the induced current is proportional to the ion velocity. Thus, Eq.~\ref{eq:equations-of-motion} can be recast in the form
\begin{equation}\label{eq:i/v-relation}
V_{ i} = L_{ i}\,\frac{dI}{dt}+\frac{1}{C_{ i}}\int{I\,dt}\;,\;i=1,\,2\;.
\end{equation}
We see that on application of an external potential $V_{ i}$ to the floating wire, the $I-V$ response
in the presence of an oscillating trapped ion is equivalent to the response of a series LC circuit. The equivalent
inductance and capacitance of the ion are
\begin{eqnarray}\label{eq:L-and-C}
&L_{i}=\frac{1}{\beta_{i}^2}\frac{m\,H^2}{e^2},\;\;\;
&C_{i}=\frac{1}{\omega_{i}^2\,L_{i}}\;,\;i=1,\,2\;,
\end{eqnarray}
with the geometry parameter, $\beta_{i}$, given by
\begin{equation}\label{eq:beta}
\beta_{i}=\frac{2\,H^2}{\alpha\,(H^2-h_{i}^2)}\;,\;i=1,\,2\;
\end{equation}
We note here that typically the inductance of a single trapped ion is \mbox{counter--intuitively large}, while the capacitance is rather small. The values corresponding to the setup parameters mentioned earlier are: $L_{1,2}\approx6\times10^4\rm H$, and $C_{1,2}\approx4\times 10^{-19}\,\rm F$.
\begin{figure}[!b]
\centering
\includegraphics[scale=0.7]{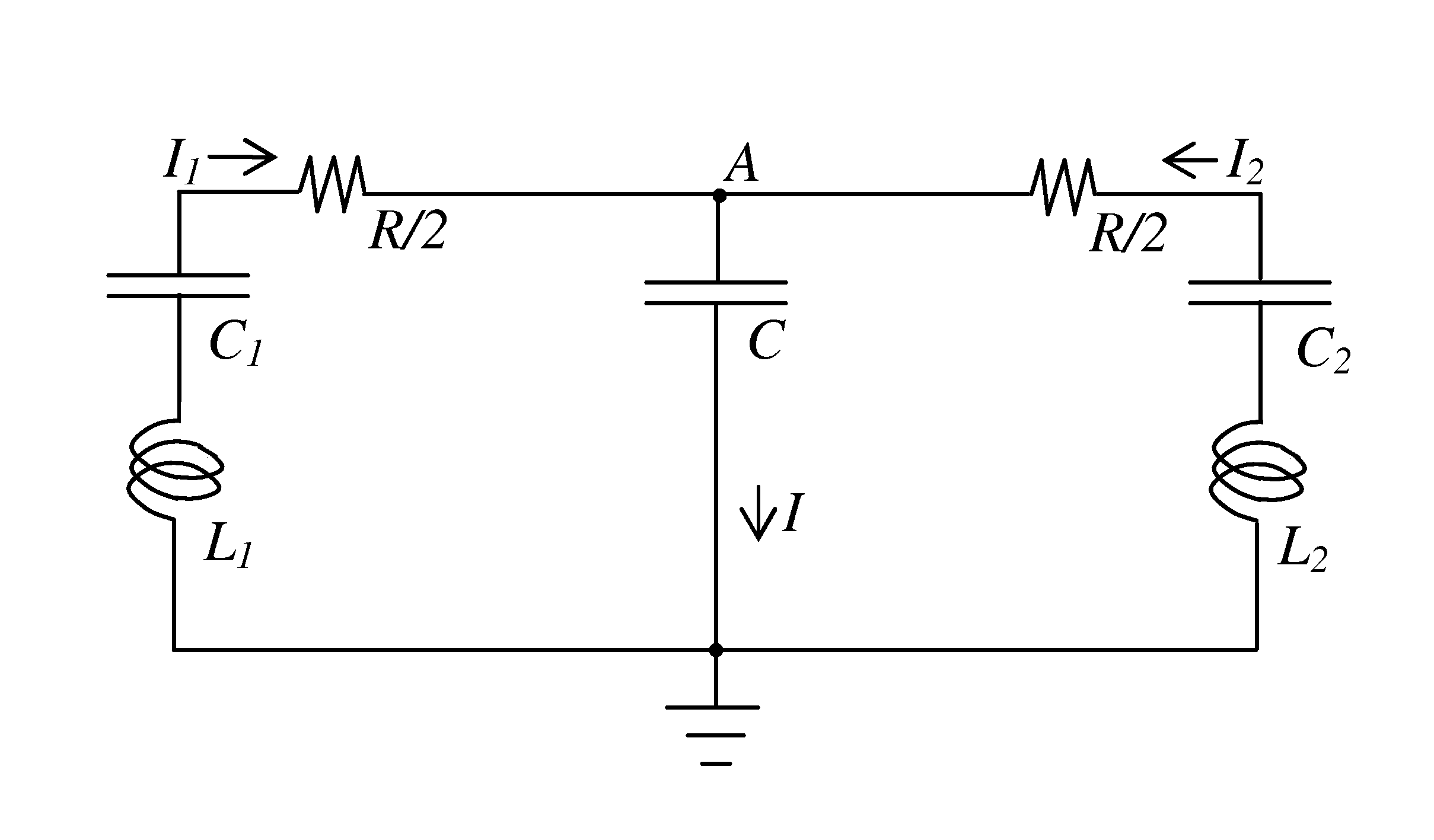}
\caption{ Equivalent circuit of two ions, each with inductance $L_i$ and capacitance $C_i$, coupled by a wire
with capacitance $C$ to ground. The wire has ohmic resistance $R$. The current $I_{i},\;i=1,\,2$ in
each branch of the circuit corresponds to the velocity of the corresponding ion.}
\label{fig:circuit}
\end{figure}
The equivalent circuit of two trapped ions coupled through a wire, with capacitance $C$ to ground and ohmic
resistance $R$, is shown in Fig.~\ref{fig:circuit}. For the wire dimensions mentioned above, the wire resistance is estimated to be approximately $0.6\,\rm\Omega$, resulting in a high quality factor $Q\approx6\times10^{11}$ for the ions.

It is interesting to point out that the classical solution of this circuit results in Eq.~\ref{eq:exchange-rate} for the rate of state exchange between the two ions. In addition, the exchange rate can be expressed in terms of circuit parameters as
\begin{equation}\label{eq:exchange-rate-2}
\frac{1}{t_{\rm ex}}=2\nu\frac{C_{i}}{C}\;,
\end{equation}
where $\nu$ is the secular frequency of the ions. By expressing the exchange rate in this form, we see that it is maximized by minimizing the wire's capacitance to ground. This is evident from the equivalent circuit: a signal originating from the motion of one ion will be shorted to ground due to the wire's capacitance. This is also the reason for keeping the coupling conductor electrically floating. Any conducting path from the coupling wire to ground will reduce the portion of the signal that couples the two ions.

The circuit model also provides insight into the effect of various decoherence sources, which we discuss next.

\section{Sources of decoherence}
\label{sec:decoherence}
The goal of these experiments is to coherently couple trapped ions with a conductor. While it is not intuitively clear whether the transport of quantum information through a macroscopic wire is possible, a theoretical analysis shows that there are no known fundamental obstacles \cite{Zurita-Sanchez:2008}.

One source of decoherence is the dissipation of the induced current inside the wire. Using Eq.~\ref{eq:induced-charge}, the current induced by a single ion with a low motional quantum number will be of order $I=e\,\dot{z}\,\beta/H\sim e\,\sqrt{\hbar\omega/m}\,\beta/H$. For the parameters mentioned before, this current amounts to approximately 0.1~fA, so we expect that it takes approximately $2\times10^5\,s$ to dissipate one motional quantum on a wire resistance of $0.6\,\rm\Omega$. More important in this context, however, is the inverse process by which the ion picks up motional quanta from Johnson noise in the wire.
The Johnson noise heating power $P$ is given by:
\begin{equation}\label{eq:Johnson-noise}
 P_{\rm noise} = kT \Delta \nu\:,
\end{equation}
where $kT$ is the thermal energy and $\Delta \nu $ is the frequency bandwidth in which the ion accepts the power. The time $\tau$  in which one motional quantum of energy $E_{\rm q}=h\nu$ is generated is  given by
\begin{equation}\label{eq:Johnson-heating}
{\tau^{-1}}=\frac{P_{\rm noise}}{E_{\rm q}}=\frac{kT\Delta \nu}{h\nu}=\frac{kT}{hQ} \:,
\end{equation}
Inserting the expression $Q=R^{-1}\sqrt{L_{i}/C_{i}}$ into Eq.~\ref{eq:Johnson-heating}, we arrive at
\begin{equation}\label{eq:heating}
 {\tau^{-1}}=\frac{kTR}{h} \sqrt{\frac{C_i}{L_i}}\:,
\end{equation}
for the inverse time in which one motional quantum is acquired. For the values used above, the expected heating time from Johnson noise is $\tau \approx 0.1$~s/quantum at room temperature, which is comparable to the exchange time, $t_{ex}$. However, this can be significantly improved by cooling to liquid helium temperatures. Assuming a resistivity ratio $\rho_{300\,K}/\rho_{4\,K}\approx50$ for the wire, the time constant for Johnson noise heating is $\tau \sim 380$~s/quantum. Thus, Johnson noise is not expected to prevent the coherent transfer of quantum information at cryogenic temperatures.

However, ion trap experiments usually report heating rates higher than what would be expected from Johnson noise \cite{Leibfried2003b,Turchette2000}. The source of this noise is believed to be fluctuating patches of charges on the electrodes. Experiments hint that coating the electrode surface with contaminants has a strong influence on the observed heating \cite{Wineland:1998a,Leibfried2003b,Turchette2000}. Furthermore, recent experiments have shown that the fluctuations are thermally activated and can be suppressed many orders of magnitude by cooling the trap electrodes \cite{Deslauriers2006a,Labaziewicz:2008a,Labaziewicz:2008b}. Indeed, heating rates as small as a few phonons/s have been observed for ions trapped as close as 75~$\mu$m to the nearest electrode of a planar trap \cite{Labaziewicz:2008a,Labaziewicz:2008b}.

Finally, we consider the effects of a leakage current from the coupling wire to ground. In the equivalent circuit, Fig. \ref{fig:circuit}, this is modeled as a large resistor, $R_g>10^{13}\,\rm\Omega$, in parallel to the capacitor $C$ between node $A$ and ground~\footnote{We point out that modeling the insulating wire support by a large resistor is a rough approximation. Typically the I-V response will be nonlinear, with nontrivial frequency scaling. Factors that contribute to an observed current in an insulator can be as different as internal polarization changes in the dielectric and surface currents due to adsorbed contaminants.}. A simple analysis shows that the current decay constant due to the resistor is $4R_gC$, estimated at more than 1~s. This time constant needs to be larger than the motional coupling timescale, $t_{\rm ex}$, which can easily be satisfied for realistic values of the leakage equivalent resistance.

\section{Experimental status}
From section \ref{sec:coupling}, it is clear that the trap should be small, allowing the ions to be close to the coupling electrode. In addition, as discussed in section \ref{sec:decoherence}, operation at cryogenic temperatures is advantageous for the reduction of decoherence processes. Both these demands are being experimentally pursued.

The experimental set-up consists of a planar surface trap housed in a vacuum vessel. \Ca ions are trapped at a height of about 200~$\mu$m above a sapphire substrate coated by gold electrodes (see Fig.~\ref{fig:set-up}). The fabrication of these traps will be described elsewhere. Typical trap frequencies are on the order of a few MHz in the radial direction and 400~kHz in the axial direction.

A gold wire with diameter $2a=25~\mu$m and length $L=10~mm$ can be positioned above the trap using \mbox{nano-positioning} stages employing a \mbox{slip-stick} mechanism. Relatively small Macor pieces support the wire and electrically isolate it from a larger metallic arm that connects the \mbox{nano-positioning} stages to the wire. In constructing the coupling wire assembly, one has to bear in mind the conflicting constrains imposed: On the one hand it is essential to minimize the wire capacitance to ground. Therefore the wire length has to be kept as small as possible, and there should be insulating pieces to separate the wire from the electrically grounded metallic arm. On the other hand, the amount of insulating material near the ions should be minimized, as they will lead to unwanted stray field effects. Modeling and optimizing the compromise between these opposing constraints is difficult. In practice, one follows an empirical, \mbox{trial-and-error} approach for the design of the coupling wire assembly.

The main goal of the current set-up is to explore the interaction of single ions with the electrically floating wire. Of interest are the influence of the wire on the trap potential, heating of the motion due to the wire, the minimum achievable ion--wire distance, and the reproducibility of day-to-day trap parameters with the electrically floating wire present. For these investigations a room temperature set-up is sufficient.

In view of the smaller dimensions required for future experiments of coherent ion-wire-ion interactions, an additional cryogenic set-up is being constructed. The cooling power will be provided by an ARS Cryosystems \mbox{closed--cycle} cryostat operating in the range of $4\,\rm K$ to $300\,\rm K$. In this setup, the trap chip, coupling wire, and nano-positioning stage are all encased in a cold stage which can reach temperatures
in the above range in a controlled way.

\section{Summary}

The possibility of coupling trapped-ion motional states through a floating conductor opens promising directions in quantum information processing and coupling quantum-optic systems to solid-state systems. Analysis of the coupling mechanism shows that the miniaturization of the ion trap and coupling electrode yields significant gain in the coupling rate. Moreover, it is possible to reduce the decoherence rates to levels much below the coupling rate by working at liquid helium temperatures. Current experiments are making progress in both directions.\\\\


\begin{thebibliography}{10}

\bibitem{Bollinger1991}
J.~J. Bollinger, D.~J. Heinzen, W.~M. Itano, S.~L. Gilbert, and D.~J. Wineland.
\newblock A 303 {MH}z frequency standard based on trapped {Be}$^+$ ions.
\newblock {\em IEEE Trans.~Instr.~Meas.}, 40:126, 1991.

\bibitem{Fisk:1997}
P.~T.~H. Fisk, M.~J. Sellars, M.~A. Lawn, and C.~Coles.
\newblock Accurate measurement of the 12.6 ghz clock transition in trapped
  $^{171}${Y}b$^+$ ions.
\newblock {\em IEEE Trans.~Ultrason.~Ferroelectr.~Freq.~Control}, 44:344, 1997.

\bibitem{Leibfried2003b}
D.~Leibfried, R.~Blatt, C.~Monroe, and D.~Wineland.
\newblock Quantum dynamics of single trapped ions.
\newblock {\em Rev.~Mod.~Phys.}, 75:281, 2003.

\bibitem{Leibfried2003a}
D.~Leibfried, B.~DeMarco, V.~Meyer, D.~Lucas, M.~Barrett, J.~Britton, W.~M.
  Itano, B.~Jelenkovi{\'c}, C.~Langer, T.~Rosenband, and D.~J. Wineland.
\newblock Experimental demonstration of a robust, high-fidelity geometric two
  ion-qubit phase gate.
\newblock {\em Nature}, 422:412--415, 2003.

\bibitem{Schmidt-Kaler2003}
F.~Schmidt-Kaler, H.~H\"affner, M.~Riebe, S.~Gulde, G.~P.~T. Lancaster,
  T.~Deuschle, C.~Becher, C.~F. Roos, J.~Eschner, and R.~Blatt.
\newblock Realization of the {C}irac-{Z}oller controlled-{NOT} quantum gate.
\newblock {\em Nature}, 422:408--411, 2003.

\bibitem{Sackett2000}
C.~A. Sackett, D.~Kielpinski, B.~E. King, C.~Langer, V.~Meyer, C.~J. Myatt,
  M.~Rowe, Q.~A. Turchette, W.~M. Itano, D.~J. Wineland, and C.~Monroe.
\newblock Experimental entanglement of four particles.
\newblock {\em Nature}, 404:256--259, 2000.

\bibitem{Leibfried2005}
D.~Leibfried, E.~Knill, S.~Seidelin, J.~Britton, R.~B. Blakestad,
  J.~Chiaverini, D.~B. Hume, W.~M. Itano, J.~D. Jost, C.~Langer, R.~Ozeri,
  R.~Reichle, and D.~J. Wineland.
\newblock Creation of a six--atom ``{S}chr\"odinger cat'' state.
\newblock {\em Nature}, 438(7068):639--642, 2005.

\bibitem{Haeffner2005a}
H.~H\"affner, W.~H\"ansel, C.~F. Roos, J.~Benhelm, D.~{Chek-al-Kar},
  M.~Chwalla, T.~K\"orber, U.~D. Rapol, M.~Riebe, P.~O. Schmidt, C.~Becher,
  O.~G\"uhne, W.~D\"ur, and R.~Blatt.
\newblock Scalable multiparticle entanglement of trapped ions.
\newblock {\em Nature}, 438:643--646, 2005.

\bibitem{Riebe2004}
M.~Riebe, H.~H\"affner, C.~F. Roos, W.~H\"ansel, J.~Benhelm, G.~P~T Lancaster,
  T.~W. K\"orber, C.~Becher, F.~Schmidt-Kaler, D.~F~V James, and R.~Blatt.
\newblock Deterministic quantum teleportation with atoms.
\newblock {\em Nature}, 429:734--737, 2004.

\bibitem{Barrett2004}
M.~D. Barrett, J.~Chiaverini, T.~Schaetz, J.~Britton, W.~M. Itano, J.~D. Jost,
  E.~Knill, C.~Langer, D.~Leibfried, R.~Ozeri, and D.~J. Wineland.
\newblock Deterministic quantum teleportation of atomic qubits.
\newblock {\em Nature}, 429:737--739, 2004.

\bibitem{Chiaverini:2004a}
J.~Chiaverini, D.~Leibfried, T.~Schaetz, M.~D. Barrett, R.~B. Blakestad,
  J.~Britton, W.~M. Itano, J.~D. Jost, E.~Knill, C.~Langer, R.~Ozeri, and D.~J.
  Wineland.
\newblock Realization of quantum error correction.
\newblock {\em Nature}, 432:602--605, 2004.

\bibitem{Haeffner:2008}
H.~H\"affner, C.F. Roos, and R.~Blatt.
\newblock Quantum computing with trapped ions.
\newblock {\em Physics Reports}, 469:155, 2008.

\bibitem{Kielpinski2001}
D.~Kielpinski, V.~Meyer, M.~A. Rowe, C.~A. Sackett, W.~M. Itano, C.~Monroe, and
  D.~J. Wineland.
\newblock A decoherence-free quantum memory using trapped ions.
\newblock {\em Science}, 291(5506):1013--1015, 2001.

\bibitem{Cirac1997}
I.~Cirac, P.~Zoller, J.~Kimble, and H.~Mabuchi.
\newblock Quantum state transfer and entanglement distribution among distant
  nodes in a quantum network.
\newblock {\em Phys.~Rev.~Lett.}, 78:3221, 1997.

\bibitem{Kimble:2008}
H.~J. Kimble.
\newblock The quantum internet.
\newblock {\em Nature}, 453(7198):1023--1030, Jun 2008.

\bibitem{Moehring2007}
D.~L. Moehring, P.~Maunz, S.~Olmschenk, K.~C. Younge, D.~N. Matsukevich, L-M.
  Duan, and C.~Monroe.
\newblock {E}ntanglement of single-atom quantum bits at a distance.
\newblock {\em Nature}, 449(7158):68--71, 2007.

\bibitem{Gottesman1999}
D.~Gottesman and I.~L. Chuang.
\newblock Quantum teleportation is a universal computational primitive.
\newblock {\em Nature}, 402:390, 1999.

\bibitem{Wineland1975}
D.~J. Wineland and H.~G. Dehmelt.
\newblock Principles of the stored ion calorimeter.
\newblock {\em J.~App.~Phys.}, 46:919, 1975.

\bibitem{Heinzen1990}
D.J. Heinzen and D.J. Wineland.
\newblock Quantum-limited cooling and detection of radio-frequency oscillations
  by laser-cooled ions.
\newblock {\em Phys.~Rev.~A}, 42:2977, 1990.

\bibitem{Tian2004a}
L.~Tian, P.~Rabl, R.~Blatt, and P.~Zoller.
\newblock {I}nterfacing quantum-optical and solid-state qubits.
\newblock {\em Phys.~Rev.~Lett.}, 92(24):247902, 2004.

\bibitem{Tian2005}
L.~Tian, R.~Blatt, and P.~Zoller.
\newblock Scalable ion trap quantum computing without moving ions.
\newblock {\em Eur.~Phys.~J.~D}, 32:201, 2005.

\bibitem{Soerensen2004}
Anders~S S{\o}rensen, Caspar~H van~der Wal, Lilian~I Childress, and Mikhail~D
  Lukin.
\newblock {C}apacitive coupling of atomic systems to mesoscopic conductors.
\newblock {\em Phys.~Rev.~Lett.}, 92(6):063601, 2004.

\bibitem{Stahl2005}
S.~Stahl, F.~Galve, J.~Alonso, S.~Djekic, W.~Quint, T.~Valenzuela,
  J.~Verd\'{u}, M.~Vogel, and G.~Werth.
\newblock A planar penning trap.
\newblock {\em Eur.~Phys.~J.~D}, 32:139, 2005.

\bibitem{Marzoli:2008}
I.~Marzoli, P.~Tombesi, G.~Ciaramicoli, G.~Werth, P.~Bushev, S.~Stahl,
  F.~Schmidt-Kaler, M.~Hellwig, C.~Henkel, G.~Marx, I.~Jex, E.~Stachowska,
  G.~Szawiola, and A.~Walaszyk.
\newblock Experimental and theoretical challenges for the trapped electron
  quantum computer.
\newblock {\em arXiv:0810.4408}, 2008.

\bibitem{Shockley:1938}
W.~Shockley.
\newblock Currents to conductors induced by a moving point charge.
\newblock {\em J.~App.~Phys.}, 9:635, 1938.

\bibitem{Jackson:1998}
J.~D. Jackson.
\newblock {\em Classical Electrodynamics, 3rd ed.}
\newblock J. Wiley \& Sond, 1998.

\bibitem{Estes:1968}
Lee~E. Estes, Thomas~H. Keil, and Lorenzo~M. Narducci.
\newblock Quantum-mechanical description of two coupled harmonic oscillators.
\newblock {\em Physical Review}, 175(1):286--299, Nov 1968.

\bibitem{Portes:2008}
S.~B.~Duarte D.~Portes, H.~Rodrigues and B.~Baseia.
\newblock Quantum states transfer between coupled fields.
\newblock {\em The European Physical Journal D}, 48(1):145--149, jun 2008.

\bibitem{Zurita-Sanchez:2008}
J.~R. Zurita-S\'anchez and C.~Henkel.
\newblock Wiring up single electron traps to perform quantum gates.
\newblock {\em New J.~Phys.}, 10:083021, 2008.

\bibitem{Turchette2000}
Q.~A. Turchette, Kielpinski, B.~E. King, D.~Leibfried, D.~M. Meekhof, C.~J.
  Myatt, M.~A. Rowe, C.~A. Sackett, C.~S. Wood, W.~M. Itano, C.~Monroe, and
  D.~J. Wineland.
\newblock Heating of trapped ions from the quantum ground state.
\newblock {\em Phys.~Rev.~A}, 61:063418, 2000.

\bibitem{Wineland:1998a}
D.~J. Wineland, C.~Monroe, W.~M. Itano, D.~Leibfried, B.~E. King, and D.~M.
  Meekhof.
\newblock Experimental issues in coherent quantum-state manipulation of trapped
  atomic ions.
\newblock {\em J.~Res.~Natl.~Inst.~Stand.~Technol.}, 103(3):259--328, 1998.

\bibitem{Deslauriers2006a}
L.~Deslauriers, S.~Olmschenk, D.~Stick, W.~K. Hensinger, J.~Sterk, and
  C.~Monroe.
\newblock {S}caling and suppression of anomalous heating in ion traps.
\newblock {\em Phys.~Rev.~Lett.}, 97(10):103007, 2006.

\bibitem{Labaziewicz:2008a}
J.~Labaziewicz, Y.~Ge, P.~Antohi, D.~Leibrandt, K.R. Brown, and I.L. Chuang.
\newblock Suppression of heating rates in cryogenic surface-electrode ion
  traps.
\newblock {\em Phys.~Rev.~Lett.}, 100:013001, 2008.

\bibitem{Labaziewicz:2008b}
Jaroslaw Labaziewicz, Yufei Ge, David Leibrandt, Shannon~X. Wang, Ruth Shewmon,
  and Isaac~L. Chuang.
\newblock Temperature dependence of electric field noise above gold surfaces.
\newblock {\em Phys. Rev. Lett.}, 101:180602, 2008.

\end{thebibliography}
\end{document}